\documentstyle[12pt]{article}
\newtheorem{th}{\large\sc Theorem}[section]
\newtheorem{pro}{\large\sc Proposition}[section]

 \newfont{\got}{eufm10 scaled \magstep1}
\newfont{\set}{msbm10 scaled \magstep1}

\begin{document}
\title{ \sc Two--Dimensional Integrable Systems\\ and \\Self--Dual
Yang--Mills Equations\thanks{Partially
 supported by CICYT proyecto PB89-0133} } \author{\large F.Guil and
{M.Ma\~nas}\thanks{Permanent address:
The Mathematical Institute, Oxford University, 24--29 St.Giles' Oxford OX1 3LB,
United Kingdom}\\ Departamento de F\'{\i}sica
Te\'orica \\
Universidad Complutense\\28040-Madrid,
Spain} \date{\,}

\maketitle
\begin{abstract}
The relation between two--dimensional integrable systems
and four--dimen\-sional self--dual Yang--Mills equations is considered.  Within
the
twistor description and the zero--curvature representation
a method is given to associate self--dual Yang-Mills
connections with integrable systems of the
 Korteweg--de Vries and
non--linear Schr\"odinger type or principal chiral models.
 Examples of self--dual connections
are constructed that
as points in the moduli do not have two independent conformal symmetries.

 \end{abstract}

\section{Introduction}

As a system of partial differential equations,
the self--dual Yang--Mills (SDYM) equations are invariant
under the action of the group of conformal
transformations acting on the four space--time coordinates.
It is well known that invariant solutions by the action
of a subgroup with two conformal generators satisfy a differential
equation in two--variables since each one--dimensional
subgroup reduces by one the number of independent
variables. These solutions are called self--similarity solutions
of the original equations.
It has been observed that this procedure allows
one to describe the corresponding invariant solutions in terms
of a two--dimensional integrable system. This is the
case for the principal chiral model$^{14,15,16}$ (for the classical euclidean
$O(1,3)$ non--linear
$\sigma$--model this was already noticed
in Ref.\cite{p}) as well as for the Korteweg--de Vries (KdV)
and Nonlinear Schr\"odinger (NLS) equations.$^{6,7}$ A review of these results
can be found in
Ref.\cite{ac}. It is a nontrivial aspect of the reduction problem
to describe the resulting two--dimensional equations.
For the principal chiral model this
 description is based in
the formulation of SDYM equations in terms of the Yang matrix.
An alternative approach consists in describing the SDYM equations as a
compatibility
condition for two linear differential equations, this
allows one
to identify the reduced solutions in the KdV and NLS cases.

It seems, however, that any reasonable relation between
two integrable equations should be
 based in their definitions as
integrable systems. The SDYM equations describe a connection
for a bundle over the Grassmannian of two--dimensional
subspaces of the twistor space.
Integrability for a SDYM connection means that its curvature
vanishes on certain two--planes in the tangent space of the
Grassmannian. As proved by Ward,$^{13,17}$ this allows
one to characterize SDYM connections in terms of the
spliting problem for a transition function in
a holomorphic
bundle over the Riemann sphere, i.e. the trivialization of
the bundle.
For two--dimensional integrable systems the situation is
quite analogous. An important feature of equations
such as the KdV, NLS or the principal chiral model consists
 in the possibility of constructing solutions through a
factorization problem in the circle. This is equivalent to
the spliting problem for the SDYM case if we choose
the transition function having the form required by the
factorization problem of a two--dimensional integrable
system.

It is the main result of this paper the construction
of a map taking arbitrary solutions of a two--dimensional
integrable system into solutions of the SDYM equations.
This map comes from the zero--curvature formulation
of a two--dimensional integrable system and represents
an extension of the correspondence derived from the
factorization problem. A generic property of the relation
for a principal chiral model, the KdV, and NLS equations
is the appearence of arbitrary functions. In particular, this generalizes
the self--similarity solutions under a group generated by
two translations. Section 2 is devoted to the analysis
of these properties. The analysis
of the correspondence for a Yang matrix is given in
Section 3 where we study its relation with a chiral field.
As a special case one gets Ward's construction$^{13,14,15}$
we
mentioned
before. Finally in Section 4 we present examples of
SDYM connections
derived from two--dimensional integrable systems
that can not be obtained by symmetry reduction by two
conformal generators.

Let us observe that the present construction does not
exhaust all possible relations between SD equations
and integrable systems. This is the case for
 the Nahm's equations that
can be associated to SDYM equations in an alternative way to
that followed in this paper.$^{5}$

\section{Integrable systems and self--dual
Yang--Mills equations} \,\hspace{.7cm}The
compatibility condition for the linear system of two
first--order differential equations on the vector
$\psi$, \begin{eqnarray*}
\psi_x&=&U(\lambda)\psi\\
\psi_t&=&V(\lambda)\psi,
\end{eqnarray*}
implies for the two matrix functions $U(x,t,\lambda),\
V(x,t,\lambda)$, depending on a complex variable
$\lambda$, the nonlinear equation
\begin{equation}
U_t-V_x+[U,V]=0.
\label{zc}
\end{equation}
Under an appropriate choice of the functions
$U(\lambda)$ and $V(\lambda)$ one can represent
through (\ref{zc}) a wide class of non--linear partial
differerential equations in the variables $x$ and $t$
known as integrable systems.$^3$ Concrete
examples for these functions that we shall consider
below are as follows.

Let $U$ and $V$ be given by the expressions
\begin{eqnarray*}
U(\lambda)&:=&\lambda U_1+U_0, \\
V(\lambda)&:=&\lambda^2 U_1+ \lambda U_0+V_0,
\end{eqnarray*}
then Eq.(\ref{zc}) implies for the coefficients $U_i$
and $V_0$ the relations
\begin{eqnarray*}
&&U_{0,t}-V_{0,x}+[U_0,V_0]=0,\\
&&U_{1,t}-U_{0,x}+[U_1,V_0]=0,\\
&&U_{1,x}=0.
\end{eqnarray*}
In particular if we take
\[U(\lambda)=\left(\begin{array}{cc}0&1\\
\lambda -u&0\end{array} \right),\ \
V(\lambda)=\left(\begin{array}{cc}-\frac{1}{4}u_x&\lambda+
\frac{1}{2}u\\
\lambda^2-\frac{1}{2}u\lambda-\frac{1}{4}
(u_{xx}+2u^2)&\frac{1}{4}u_x\end{array}\right),\nonumber
\]
we obtain the Korteweg--de Vries (KdV) equation$^8$
 for
the scalar function $u(x,t)$
\[
4u_t=u_{xxx}+6uu_x.
\nonumber
\]

Letting now $U$ and $V$ be  given by
\[
U(\lambda)=\left(\begin{array}{cc}i\lambda&p\\
-p^{\ast}&-i\lambda\end{array}
\right),\ \
V(\lambda)=\left(\begin{array}{cc}i\lambda^2-\frac{i}{2}|p|^2
&\lambda p-\frac{i}{2}p_x\\
-\lambda p^{\ast}-\frac{i}{2}p^{\ast}_x
&-i\lambda^2+\frac{i}{2}|p|^2\end{array}\right),
\nonumber
\]
the system above results in the nonlinear
Sch\"odinger (NLS) equation for the complex scalar field $p(x,t)$
\[
ip_t=\frac{1}{2}p_{xx}+|p|^2p.
\]

The equations of a principal chiral field for
functions $u(x,y),\ v(x,y)$ with values in a Lie
algebra {\got g }
\begin{equation}
u_y+\frac{1}{2}[u,v]=0,\ \ \ v_x-\frac{1}{2}[u,v]=0
\label{quiral}
\end{equation}
can  equally be represented with the aid of (\ref{zc})
if we choose in this case
\[
U(\lambda)=-\frac{u}{\lambda -1},\ \ \ V(\lambda)=
\frac{v}{\lambda+1}
\]
with the substitution $t\rightarrow y$.$^{19}$

The preceding formulas can be conveniently
represented as the zero curvature condition for a
connection
\[
\begin{array}{ccccc}
\omega &:& \mbox{\set C}&\rightarrow &
\bigwedge^1(M,\mbox{\got g})\\
& &\lambda&\mapsto &\omega(\lambda),
\end{array}
\]
where $\lambda$ is a spectral parameter and $\omega(\lambda)$ is
a  {\got g}--valued 1--form over the two dimensional
time manifold $M$ having as local coordinates the variables $x,t$ (or $x$ and
$y$).
 In terms of local coordinates
 $x,y$ (or $x,t$) of the manifold $M$ we can write the connection as
\[
\omega(\lambda)=U(\lambda)dx+V(\lambda)dy,
\]
for which the vanishing of the curvature
\[
\Omega(\lambda):=d\omega(\lambda)-
\frac{1}{2}[\omega(\lambda),\omega(\lambda)]
\]
is equivalent to the relation (\ref{zc}).

Self--dual Yang--Mills (SDYM) equations constitute
another important example of non--linear
partial differential equations which are also integrable.
Now, we shall be dealing with a four dimensional system
instead of the two--dimensional cases previously
analyzed. If $z=(z^{AA^{\prime}}), A=0,1, A^{\prime}=
0^{\prime}, 1^{\prime}$ represents a $2\times 2$
complex matrix describing the local coordinates of  a
four dimensional complex manifold and $\Phi =
\Phi_{AA^{\prime}}dz^{AA^{\prime}}$ is a {\got
g}--valued connection described by the set of functions
$\Phi_{AA^{\prime}}(z)$, then the curvature 2--form
$F=d\Phi-\frac{1}{2}[\Phi,\Phi]$ can be written as
\[
F=F_{AA^{\prime}BB^{\prime}}dz^{AA^{\prime}}\wedge
dz^{BB^{\prime}}.
\]
A self--duality condition$^{17}$ for the Yang--Mills fields
$F_{AA^{\prime}BB^{\prime}}$ follows from the
decomposition
$F_{AA^{\prime}BB^{\prime}}=\varphi_{AB}
\varepsilon_{A^{\prime}B^{\prime}}+
\varphi_{A^{\prime}B^{\prime}}\varepsilon_{AB}$
where $\varepsilon$ is the  Levi--Civita tensor. Thus,
the connection $\Phi$ is  self--dual (SD) if \[
\varphi_{A^{\prime}B^{\prime}}=0.
\]

Given a SD connection $\Phi$ any gauge transformation
$\Phi\rightarrow \Phi^g=dg\cdot g^{-1}+\mbox{Ad}g\ \Phi$
gives again a    SD connection. A conformal transformation
of coordinates $z\mapsto z=(A\cdot z+B)\cdot(C\cdot z+D)^{-1}$,
with $\left(\begin{array}{cc}A&B\\
C&D\end{array}\right)\in SL(4,\mbox{\set C})$,
generates a local diffeomorphism $\phi$ such that $\phi^{\ast}\Phi$  is
also a SD connection. We say that two SD connections are equivalent if there
exists a gauge and/or a conformal transformation taking one into the other.
When we introduce certain topological requirements this space of equivalence
classes
of SD connections is called the moduli space.

The structure of this set of non--linear partial
differential equations and its relation with
the two--dimensional integrable systems presented
before are best understood through the geometry of a
complex vector bundle over the Riemann sphere$^{13,17}$ as we shall explain
briefly.

Let us denote by $T$ a four--dimensional complex
linear space, the twistor space,$^9$ and
consider
its decomposition as a
direct sum of a pair of two--dimensional subspaces $S$ and
$S^{\prime}$, $T=S\oplus S^{\prime}$. We associate to
this pair of subspaces a coordinate chart of the
Grassmannian manifold of two--dimensional subspaces
of $T$ as follows. If $V$ is a subspace of $T$ that does not
intersect $S^{\prime}$, we assign to $V$ the linear
map $z:S^{\prime}\rightarrow S$ having as graph $V$.
 Then, one is allowed to identify the coordinates
$z^{AA^{\prime}}$ on which the connection
$\Phi=\Phi_{AA^{\prime}}dz^{AA^{\prime}}$
depends with the coordinate functions of these points
in the Grassmannian. If a point in $T$ has coordinates
$(x^A,x_{A^{\prime}})$ adapted to the representation
$T=S\oplus S^{\prime}$, the linear subspace $V$
determined by the linear transformation $z$ is
characterized by the relations
$x^A=z^{AA^{\prime}}x_{A^{\prime}}$.

Let $\lambda^{A^{\prime}}$ denote the coordinates of a
point in the dual   of  $S^{\prime}$ and define the
vector fields on the Grassmannian
\[
\partial_A:=\partial_{AA^{\prime}}\lambda^{A^{\prime}},
\]
given in terms of the local coordinates $z$  by the
differential operators
$\partial_{AA^{\prime}}:=\partial /\partial
z^{AA^{\prime}}$ that generate the tangent space. The
SD equations for the connection $\Phi$ are then
equivalent to the requirement that $F$ vanishes over
the two--plane $\mbox{\set C}\{\partial_A\}_{A=0,1}$
\begin{equation}
F(\partial_A,\partial_B)=0.
\label{zcsd}
\end{equation}
One has
\begin{eqnarray*}
F(\partial_A,\partial_B)&=&\lambda^{A^{\prime}}
\lambda^{B^{\prime}}F(\partial_{AA^{\prime}},
\partial_{BB^{\prime}})=\lambda^{A^{\prime}}
\lambda^{B^{\prime}}(\varphi_{AB}
\varepsilon_{A^{\prime}B^{\prime}}+\varphi_{A^{\prime}
B^{\prime}}\varepsilon_{AB})\\
&=&\lambda^{A^{\prime}}
\lambda^{B^{\prime}}\varphi_{A^{\prime}
B^{\prime}}\varepsilon_{AB}
\end{eqnarray*}
from which our assertion follows. The connection
$\Phi$ satisfies the equation
\[
[\partial_A-\Phi_A,\partial_B-\Phi_B]=\partial_B\Phi_A-
\partial_A\Phi_B+[\Phi_A,\Phi_B]=0
\]
that we get from (\ref{zcsd}) with
$\Phi_A:=\Phi(\partial_A)=\Phi_{AA^{\prime}}
\lambda^{A^{\prime}}$ and the SD equations are
\begin{equation}
F_{00^{\prime}10^{\prime}}=
F_{00^{\prime}11^{\prime}}+F_{01^{\prime}10^{\prime}}=
F_{01^{\prime}11^{\prime}}=0.\label{sde}
\end{equation}
All relations involving the coordinates
$\lambda^{A^{\prime}}$ are homogeneous in
these variables, they are therefore well defined on the
proyective $\lambda^{A^{\prime}}$ plane. With the
standard covering by the two charts $\mbox{\set C}_+,
\mbox{\set C}_-$
\begin{eqnarray*}
\mbox{\set C}_+&:=&\{
\lambda^{1^{\prime}}/\lambda^{0^{\prime}},
\lambda^{0^{\prime}}\neq 0\}\\
\mbox{\set C}_-&:=&\{
\lambda^{0^{\prime}}/\lambda^{1^{\prime}},
\lambda^{1^{\prime}}\neq 0\}
\end{eqnarray*}
the induced equations are
\begin{eqnarray*}
\left[\partial_A^+-\Phi_A^+,\partial_B^+-\Phi_B^+\right]=0,
& &\partial_A^+:=\partial_{A0^{\prime}}+
\lambda\partial_{A1^{\prime}},\ \
\Phi_A^+:=\Phi_{A0^{\prime}}+
\lambda\Phi_{A1^{\prime}},\ \ \lambda\in
\mbox{\set C}_+,\\
\left[\partial_A^--\Phi_A^-,
\partial_B^--\Phi_B^-\right]
=0,
&& \partial_A^-:=\lambda\partial_{A0^{\prime}}+
\partial_{A1^{\prime}}, \  \
\Phi_A^-:=\lambda\Phi_{A0^{\prime}}+
\Phi_{A1^{\prime}}, \ \ \lambda\in
\mbox{\set C}_-.
\end{eqnarray*}
 Then, we have  trivializations
$\psi_{\pm}:\mbox{\set C}_{\pm}\rightarrow G$ defined by
\[
\partial_A^+\psi_+=\Phi_A^+\psi_+,\ \partial_A^-\psi_-=\Phi_A^-\psi_-,
\]
that on $\mbox{\set C}_+\cap\mbox{\set C}_-$ solve the
splitting problem for the transition function $\psi$,
$\psi_+=\psi\cdot\psi_-$.$^{13,17,12}$ This was just the
situation for the two--dimensional integrable systems
considered before
and tells us
 about the possibility of describing SD connections by means of
integrable two--dimensional non--linear partial
differential equations.
This is precisely the case if we let
$\omega(\lambda)$  be defined on the tangent vectors
$\partial_{AA^{\prime}}$, equivalently the coordinates
$x,t$ (or $x,y$) are functions of $z$ and $\lambda$
belongs to $\mbox{\set C}_+$ or $\mbox{\set C}_-$.
For suitable functions $x(z), t(z)$ we define the SD
connection $\Phi$ associated to $\omega$ by the
 relations
\begin{equation}
\Phi(\partial_A)=\omega(\partial_A).
\label{sdis}
\end{equation}
The zero--curvature condition for $\omega$,
$d\omega-1/2[\omega,\omega]=0$, on the vectors
$\partial_A,\partial_B$ implies the SD equations for
$\Phi$:
\begin{eqnarray*}
0&=&(d\omega-\frac{1}{2}[\omega,\omega])
(\partial_A,\partial_B)=[\partial_A-\omega(\partial_A),
\partial_B-\omega(\partial_B)]\\ &=&
[\partial_A-\Phi(\partial_A),
\partial_B-\Phi(\partial_B)]=F(\partial_A,\partial_B).
\end{eqnarray*}
The 1--form $\omega$ induces  a self--dual connection if its
contraction with $\partial_A$ depends on $\lambda$
as $\Phi(\partial_A)$ does.
Explicit expressions for $\omega$ are obtained in each
case by imposing the condition (\ref{sdis}) upon the
$\lambda$--dependent 1--form
$\omega$.   Now, we are in a position to formulate the relation
between self--dual connections and integrable systems of KdV and NLS type.

\begin{th}
Let $\omega=U(\lambda)dx+V(\lambda)dt$ be the
zero--curvature 1--form
\[
\omega(\lambda)=(\lambda U_1+U_0)dx+
(\lambda^2 U_1+\lambda U_0+V_0)dt.
\]
Then, there exists   a self--dual
connection $\Phi$ asociated to $\omega(\lambda)$
in $\mbox{\set C}_+$ if
\begin{eqnarray*}
t(z)&=&m(z^{00^{\prime}}, z^{10^{\prime}})\\
x(z)&=&-z^{01^{\prime}}\partial_{00^{\prime}}m(z)-
z^{11^{\prime}}\partial_{10^{\prime}}m(z)+
n(z^{00^{\prime}},
z^{10^{\prime}}) \end{eqnarray*}
for arbitrary functions $m$ and $n$. The coefficients of the
self--dual
connection are given by
\begin{eqnarray*}
\Phi_{A0^{\prime}}&=&U_0\partial_{A0^{\prime}}x+V_0
\partial_{A0^{\prime}}t\\
\Phi_{A1^{\prime}}&=&U_1\partial_{A0^{\prime}}x.
\end{eqnarray*}
\end{th}
{\bf Proof:}
On $\mbox{\set C}_+$ we have
$\partial_A^+:=\partial_{A0^{\prime}}+
\lambda\partial_{A1^{\prime}}$ and
\begin{eqnarray*}
\omega(\partial_A^+)&=&U(\lambda)\partial_A^+x+V(\lambda)
\partial_A^+t\\
\ &=&(\lambda U_1+U_0)(\partial_{A0^{\prime}}x+
\lambda\partial_{A1^{\prime}}x)+
(\lambda^2 U_1+\lambda U_0+V_0)
(\partial_{A0^{\prime}}t+
\lambda\partial_{A1^{\prime}}t).
\end{eqnarray*}
Then, Eq.(\ref{sdis}) implies that
 the coefficients of
$\lambda^3$ and $\lambda^2$
must vanish and this gives the desired result.$\Box$

The second class of SD connections arising from
two--dimensional integrable systems we shall
consider is related to the equations of the principal
chiral field. The proof of the following theorem
reproduces the preceding one.

\begin{th}
Let the 1-form
\[
\omega(\lambda):=-\frac{u}{\lambda -1}dx+
\frac{v}{\lambda +1}dy
\]
have zero--curvature for $|\lambda| <1$ in $\mbox{\set
C}_-$ and $x,y$ be functions of the form
\begin{eqnarray*}
x&=&x(z^{00^{\prime}}-z^{01^{\prime}},z^{10^{\prime}}-z^{11^{\prime}})\\
y&=&y(z^{00^{\prime}}+z^{01^{\prime}},z^{10^{\prime}}+z^{11^{\prime}}),
\end{eqnarray*}
then the associated connection
 $\Phi(\partial_A)=\omega(\partial_A)$
 with components given by
\[
\Phi_{A0^{\prime}}=0,\ \
\Phi_{A1^{\prime}}=u\partial_{A1^{\prime}}x+
v\partial_{A1^{\prime}}y
\]
is self--dual.
\end{th}

We observe that the coordinates $x,y$ on which a
principal chiral field depends determine a SD
connection if they are defined on the image under $z$
of the proyective lines $\mbox{\set C}(0,0,1,-1)$ and
$\mbox{\set C}(0,0,1,1)$
 respectively.

Integrable systems of the type considered in {\sc Theorem
 2.1} form a large family. Let us mention the modifications
of KdV and NLS, the Fordy--Kulish NLS equations in homogeneous
spaces, or the Burgers equation.
Similarly, the principal chiral model
 contains several  integrable systems as those derived from
 the theory of harmonic maps, not only in Lie groups,
 but in general
in Grassmannians.$^{11}$ Also it has a number of  reductions,
such as $\sigma$--models, the Gross--Neveu
model and others.$^{20}$
Even in the simplest case
 $\mbox{\got g}=\mbox{\got sl}(2,\mbox{\set C})$ this model has
 interesting reductions. Let us mention the sinh--Gordon equation,
see Ref.\cite{u} for an analysis of the relations of this equation
with SDYM and harmonic maps,
the  massive Thirring model, and the self--induced
transparency equations. As examples,
 we shall write down explicitly the details for
the two
first mentioned integrable systems: the sinh--Gordon equation
and the massive Thirring model.
Consider the action
of the homographic transformation
 $\lambda\rightarrow\frac{\lambda-1}{\lambda+1}$
on the zero--curvature 1--form $\omega $ of
 {\sc Theorem 2.2},
the result is
 the zero--curvature representation for harmonic maps used in
Ref.\cite{Uhl}. Now an arbitrary gauge transformation
 gives a new zero--curvature 1--form
\[
\omega(\lambda)=(\lambda L_1+L_0)dx+(M_0+\lambda^{-1}M_1)dy.
\]
In the chart obtained by a left Lorentz transformation of coordinates
\[
z\rightarrow\frac{1}{\sqrt{2}}
\left(
\begin{array}{cc} 1&-1\\1&1\end{array}
\right)\cdot z
\]
we have a SD connection
\begin{eqnarray*}
\Phi_{A0^{\prime}}&=&L_0\partial_{A0^{\prime}}x-M_1\partial_{A1^{\prime}}
y,\\
\Phi_{A1^{\prime}}&=&M_0\partial_{A1^{\prime}}y-L_1\partial_{A0^{\prime}}
x.
\end{eqnarray*}
Let $\{E,H,F\}$ be the standard Cartan--Weyl basis of
$\mbox{\got sl}(2,\mbox{\set C})$.
For the sinh-Gordon equation we have
\begin{eqnarray*}
&&L_0=\exp(-r)E+\frac{r_x}{2}H,\ \ L_1=\exp(r)F\\
&&M_0=\exp(-r)F+\frac{r_y}{2}H,\ \ M_1=\exp(r)E,
\end{eqnarray*}
where the function $r$ satisfies the sinh--Gordon equation
\[
r_{xy}=2\sinh(2r).
\]
If we take now
\begin{eqnarray*}
&&L_0=2uE-i|u|^2H,\ \ L_1=iH-2u^{\ast}F\\
&&M_0=-2v^{\ast}F-i|v|^2H,\ \ M_1=iH+2vE,
\end{eqnarray*}
we obtain for the functions $u$ and $v$ the equations
of the massive Thirring model
\begin{eqnarray*}
iu_y&=&2u|v|^2-2v\\
iv_x&=&-2v|u|^2-2u.
\end{eqnarray*}

Conversely, if we let $x$ and $t$ be two functions of
$z$ as in {\sc Theorem 2.1} for which
\begin{equation}
\Delta:=\frac{\partial(x,t)}
{\partial(z^{00^{\prime}},z^{10^{\prime}})}=
\partial_{00^{\prime}}x\partial_{10^{\prime}}t-
\partial_{10^{\prime}}x\partial_{00^{\prime}}t
\neq 0
\label{delta}
\end{equation}
and we construct the differential 1-form
$\omega(\lambda)=U(\lambda)dx+V(\lambda)dt$, where
we define $U_0,U_1,V_0$ according to the formulas
\begin{eqnarray}
U_0&:=&\frac{1}{\Delta}(\Phi_{00^{\prime}}
\partial_{10^{\prime}}t-
\Phi_{10^{\prime}}
\partial_{00^{\prime}}t) \nonumber\\
U_1&:=&\frac{1}{\Delta}(\Phi_{00^{\prime}}
\partial_{10^{\prime}}x-
\Phi_{10^{\prime}}
\partial_{00^{\prime}}x)\nonumber\\
V_0&:=&\frac{1}{\Delta}(\Phi_{01^{\prime}}
\partial_{10^{\prime}}t-
\Phi_{11^{\prime}}
\partial_{00^{\prime}}t)\label{gor},
\end{eqnarray}
then we obtain the inversion of the formulas given in
{\sc Theorem 2.1} defining the connection $\Phi$
in terms of the coefficients $U_0,U_1,V_0$
of $\omega$.
\begin{pro}
Let $\omega(\lambda)$ the 1-form given by
$\omega(\lambda)=(\lambda U_1+U_0)dx+
(\lambda^2 U_1+\lambda U_0+V_0)dt$, where $U_0,V_0,U_1$
are defined in (\ref{gor}) and depend on $z$ through the functions
$x,t$ of {\sc Theorem 2.1}. Then the curvature
$d\omega-1/2[\omega,\omega]=0$ if
$\Phi$ is self--dual and satisfies the gauge  condition
\[
\Phi_{01^{\prime}}\partial_{10^{\prime}}x-
\Phi_{11^{\prime}}\partial_{00^{\prime}}x=0.
\]

\end{pro}

{\bf Proof:}
{}From the expressions for $\Phi_{AA^{\prime}}$ in
 {\sc Theorem 2.1} and the relations satisfied by $x$
and $t$, we get
the following expressions for
 the curvature components,
\begin{eqnarray*}
F_{00^{\prime}10^{\prime}}&=&\Delta
\ (U_{0,t}-V_{0,x}+[U_0,V_0])\\
F_{01^{\prime}11^{\prime}}&=&\Delta\  U_{1,x}\\
F_{00^{\prime}11^{\prime}}
+F_{01^{\prime}10^{\prime}}&=&\Delta\ (U_{1,t}-U_{0,x}
-[V_0,U_1]).
\end{eqnarray*}
The curvature
$\Omega=d\omega-1/2[\omega,\omega]$  becomes
\[
\Omega(\lambda)=dx\wedge
dt\ \left\{\lambda^2U_{1,x}+\lambda
(U_{0,x}-U_{1,t}-[U_1,V_0])+V_{0,x}-U_{0,t}-[U_0,V_0]
\right\}
\]
that vanishes upon the SD equations.$\Box$

For the principal chiral fields  one has,

\begin{pro}

Let $x$ and $y$ be as in {\sc Theorem 2.2} and define
\begin{eqnarray*}
u&:=&\frac{1}{\Delta}(\Phi_{01^{\prime}}
\partial_{11^{\prime}}y-\Phi_{11^{\prime}}
\partial_{01^{\prime}}y)\\
v&:=&-\frac{1}{\Delta}(\Phi_{01^{\prime}}
\partial_{11^{\prime}}x-\Phi_{11^{\prime}}
\partial_{01^{\prime}}x)
\end{eqnarray*}
where
\[
\Delta=\frac{\partial (x,y)}{\partial (z^{01^{\prime}},
z^{11^{\prime}})}.
\]
Suppose that $\Phi$ is chosen such that $u$ and $v$ depends on $z$
through the functions $x$ and $y$ defined
 in {\sc Theorem 2.2}, then
$u$ and $v$ are solutions
of the equations of a pincipal chiral field if
$\Phi$ is self--dual an satisfies the gauge condition
 $\Phi_{A0^{\prime}}=0$.
\end{pro}

{\bf Proof:}
The choice made for $u,v,x$ and $y$ together with the
gauge condition  $\Phi_{A0^{\prime}}=0$  imply the
relations
\begin{eqnarray*}
F_{00^{\prime}10^{\prime}}&=&0\\
F_{01^{\prime}11^{\prime}}&=&\Delta\ (u_y-v_x+[u,v])\\
F_{00^{\prime}11^{\prime}}
+F_{01^{\prime}10^{\prime}}&=&\Delta\ (u_y+v_x)
\end{eqnarray*}
and the result follows.$\Box$

All the  2D integrable equations cited above
has a common feature, their zero--curvature
formulation has a rational dependence in the spectral parameter $\lambda$.
But there exists 2D integrable systems  with an elliptic dependence in the
spectral
parameter such as the Landau-Lifshitz and
the Krichever--Novikov
equations. It is an open question whether they
 are related to the standard SDYM
equations  or  there exists
 an elliptic deformed version of the SDYM equations.

\section{The Yang matrix and chiral fields}
\setcounter{equation}{0}
\setcounter{th}{0}
 The equation of a principal field (\ref{quiral}) can be
equally written as a pair of conditions
\begin{eqnarray*}
&&u_y-v_x+[u,v]=0,\\
&&u_y+v_x=0.
\end{eqnarray*}
The first of them represents the vanishing of the
curvature for the connection $udx+vdy$ and this allows
one to introduce the chiral field $s$ related to its
currents $u$ and $v$ by the formulas
\[
u=s_x\cdot s^{-1},\ \ v=s_y\cdot s^{-1}.
\]
Here $s$ represents a function of $x$ and $y$ with
values in the Lie group under consideration. Then, upon
 substitution in the second equation, we get for $s$
the new condition
\[
(s_x\cdot s^{-1})_y+(s_y\cdot s^{-1})_x=0.
\]
The situation we have just described has a precise
analogue for the self--duality equations previously
considered. As follows from {\sc Theorem 2.2}, one can
construct a SD connection in  terms of a principal
chiral field with the gauge condition
\[
\Phi_{A0^{\prime}}=0.
\]
This type of SD connections can be conveniently
represented by means of a function $J(z)$,
the Yang matrix,  with values in the corresponding Lie
group.$^{10,18}$ The SD equations $F_{00^{\prime}10^{\prime}}=
F_{01^{\prime}11^{\prime}}=0$ in terms of the connection
$\Phi_{AA^{\prime}}$ are
\begin{eqnarray*}
\left[ \partial_{00^{\prime}}-\Phi_{00^{\prime}},
\partial_{10^{\prime}}-\Phi_{10^{\prime}}\right]&=&0,\nonumber\\
\left[ \partial_{01^{\prime}}-\Phi_{01^{\prime}},
\partial_{11^{\prime}}-\Phi_{11^{\prime}}\right]&=&0.
\end{eqnarray*}
This pair of zero--curvature conditions  allows us to
find functions $\varphi_{0^{\prime}}$ and
$\varphi_{1^{\prime}}$ for which
\[
\Phi_{A0^{\prime}}=
\partial_{A0^{\prime}}\varphi_{0^{\prime}}
\cdot\varphi_{0^{\prime}}^{-1},\ \
\Phi_{A1^{\prime}}=
\partial_{A1^{\prime}}\varphi_{1^{\prime}}
\cdot\varphi_{1^{\prime}}^{-1}
.\]
We can write the connection $\Phi$ as
\begin{eqnarray*}
\Phi&=&\Phi_{AA^{\prime}}dz^{AA^{\prime}}=
\partial_{A0^{\prime}}\varphi_{0^{\prime}}
\cdot\varphi_{0^{\prime}}^{-1}dz^{A0^{\prime}}+
\partial_{A1^{\prime}}\varphi_{1^{\prime}}
\cdot\varphi_{1^{\prime}}^{-1}dz^{A1^{\prime}}=\\
&=&d\varphi_{0^{\prime}}
\cdot\varphi_{0^{\prime}}^{-1}+
\mbox{Ad}\varphi_{0^{\prime}}
(\partial_{A1^{\prime}}J
\cdot J^{-1}dz^{A1^{\prime}})
\end{eqnarray*}
where we have set
\[
J:=\varphi_{0^{\prime}}^{-1}\cdot\varphi_{1^{\prime}}.
\]
Thus it appears that every SD connection is gauge
equivalent to the one given by the formulas
\[
\Phi_{A0^{\prime}}=0,\ \ \Phi_{A1^{\prime}}=
\partial_{A1^{\prime}}J\cdot J^{-1}dz^{A1^{\prime}}
\]
provided $J$ satisfies the equation
\begin{equation}
\partial_{00^{\prime}}(\partial_{11^{\prime}}
J\cdot J^{-1})
-\partial_{10^{\prime}}(\partial_{01^{\prime}}
J\cdot J^{-1})
=0,
\label{yang}
\end{equation}
which is equivalent to the SD condition $F_{00^{\prime}11^{\prime}}+
F_{01^{\prime}10^{\prime}}=0$. Moreover, from the
relation contained in {\sc Theorem 2.2} we obtain
\[
\partial_{A1^{\prime}}J\cdot J^{-1}=
s_x\cdot s^{-1}\partial_{A1^{\prime}}x+s_y\cdot s^{-1}
\partial_{A1^{\prime}}y
\]
what tell us that for a given chiral field $s(x,y)$, we
can take $J(z)=s(x(z),y(z))$
as a Yang matrix if the  coordinate functions $x,y$
are those
prescribed by {\sc Theorem 2.2}. In fact one finds the
explicit relation
\[
\partial_{00^{\prime}}(\partial_{11^{\prime}}
J\cdot J^{-1})
-\partial_{10^{\prime}}(\partial_{01^{\prime}}
J\cdot J^{-1})
=\Delta\ (
(s_x\cdot s^{-1})_y+(s_y\cdot s^{-1})_x)
\]
with
\[
\Delta=\frac{\partial (x,y)}{\partial (z^{01^{\prime}},
z^{11^{\prime}})}.
\]

\section{Reduction by symmetries}
\setcounter{equation}{0}
\setcounter{th}{0}

The conformal group represents the symmetry group for the
SDYM equations. Invariant solutions under the action of
a subgroup of rank two are described by a system of
equations
containing two independent variables instead of the
variables
$z^{AA^{\prime}}$ appearing in the original equations. Thus,
it is possible to describe SD connections possesing
two translational symmetries in terms of KdV and NLS equations.$^6$ These
connections correspond
to a real form $\mbox{\set R}^{2,2}$ of the twistor space
$T$ and the choice
$m(z)=z^{00^{\prime}}$ and
$n(z)=-z^{10^{\prime}}$ for the two functions
in the formula of {\sc Theorem 2.1}. An analogous result
for the principal chiral field$^{10,14,12}$ follows from
{\sc Theorem 2.2} when we define
$x(z)=z^{00^{\prime}}-
z^{01^{\prime}}$ and
 $y(z)=z^{10^{\prime}}+z^{11^{\prime}}$ in that case.

In this section we present examples of connections derived from
our general construction and not having two independent
conformal symmetries.

The group of transformations preserving up to a conformal factor
the symmetric bilinear form
\[
\mbox{\sf g}=\mbox{\sf g}_{AA^{\prime}BB^{\prime}}
dz^{AA^{\prime}}
dz^{BB^{\prime}}=
dz^{10^{\prime}} dz^{01^{\prime}}-dz^{11^{\prime}}
dz^{00^{\prime}}
\]
coincides with the conformal group
defined in Section 2.
If a conformal transformation has as fundamental
 vector field
$X=\sum_{AA^{\prime}}X^{AA^{\prime}}
\partial_{AA^{\prime}}$, then the coefficients
 $X^{AA^{\prime}}(z)$ are rational functions of
 $z^{AA^{\prime}}$ and are characterised by the condition
\[
\mbox{\pounds}_X \mbox{\sf g}=\Lambda\mbox{\sf g},
\]
where $\mbox{\pounds}_X$  is
the Lie derivative operator along $X$ and
$\Lambda$ represents the infinitesimal conformal factor.
In components, the condition above  reads
\[
\mbox{\sf g}_{AA^{\prime}CC^{\prime}}\partial_{BB^{\prime}}X^{CC^{\prime}}+
\mbox{\sf g}_{CC^{\prime}BB^{\prime}}\partial_{AA^{\prime}}
X^{CC^{\prime}}=
\Lambda\mbox{\sf g}_{AA^{\prime}BB^{\prime}},
\]
or in a more explicit form
\begin{eqnarray*}
&&\partial_{00^{\prime}}X^{11^{\prime}}=
\partial_{10^{\prime}}X^{01^{\prime}}=
\partial_{01^{\prime}}X^{10^{\prime}}=
\partial_{11^{\prime}}X^{00^{\prime}}=0,\\
&&\partial_{00^{\prime}}X^{01^{\prime}}=\partial_{10^{\prime}}X^{11^{\prime}},
\partial_{00^{\prime}}X^{10^{\prime}}=\partial_{01^{\prime}}X^{11^{\prime}},
\partial_{11^{\prime}}X^{01^{\prime}}=\partial_{10^{\prime}}X^{00^{\prime}},
\partial_{11^{\prime}}X^{10^{\prime}}=\partial_{01^{\prime}}X^{00^{\prime}},\\
&&\partial_{11^{\prime}}X^{11^{\prime}}+\partial_{00^{\prime}}X^{00^{\prime}}=
\partial_{01^{\prime}}X^{01^{\prime}}+\partial_{10^{\prime}}X^{10^{\prime}}=
\Lambda.
\end{eqnarray*}

Suppose that the connection $\Phi$ has a
conformal  symmetry  generated by $X$.
Then it follows$^4$
bv that $\Phi$ satisfies an equation of the form
\[
 \mbox{\pounds}_{X}(\Phi)=
dW+[ W,\Phi],
\]
for some $ W: T\rightarrow\mbox{\got g}$ that under
 a gauge transformation generated by $g$
 transforms according to
$W\mapsto  X(g)\cdot g^{-1}+\mbox{Ad}g  W$.
 For the curvature $F$ we have the condition
\[
\mbox{\pounds}_{X}F=[W, F],
\]
and for any Ad--invariant bilinear form  B
in the Lie algebra {\got g} we obtain
\begin{equation}
\mbox{\pounds}_{X}\mbox{B}(F,F)=
\mbox{B}([W,F],F)+\mbox{B}(F,[W,F])=0.\label{ecuaa}
\end{equation}
This is a necesary condition
in order to  $X$  generates a conformal symmetry of
the connection $\Phi$.

A simple computation proves that
\[
\mbox{B}(F,F)={\cal F}\Omega
\]
where $\Omega=dz^{00^{\prime}}\wedge dz^{01^{\prime}}\wedge dz^{10^{\prime}}
\wedge dz^{11^{\prime}}\in\Lambda^4(T)$
 is the standard volume form on $T$,
and therefore
\[
\mbox{\pounds}_{X}\mbox{B}(F,F)=
(X{\cal F}+{\cal F}
\mbox{div}_{\Omega}X)
\Omega,
\]
where
$\mbox{div}_{\Omega}X=\sum_{AA^{\prime}}\partial_{AA^{\prime}}
X^{AA^{\prime}}$ is the standard divergence
of a vector field that for
a conformal vector field is $\mbox{div}_{\Omega}X=2\Lambda$.
Equation (\ref{ecuaa}) for the vector field $X$ can be
written now as the equivalent condition
\begin{equation}
X{\cal F}+{\cal F}
\mbox{div}_{\Omega}X=0.
                \label{ecuab}
\end{equation}

Let us now consider the particular solution
to the principal chiral field model given by
\[
s(x,y)=\exp((x+y)A)\cdot g.
\]
Here $A$ and $g$ are constants elements
in {\got g} and $G$ respectively,
with the normalization condition
 $\mbox{B}(A,A)=1$.  For the
functions $u=s_x\cdot s^{-1}$ and $v=s_y\cdot s^{-1}$
we obtain
$u=v=A$, and the associated connection of {\sc Theorem 2.2} becomes
\begin{equation}
\Phi_{A0^{\prime}}=0,\ \Phi_{A1^{\prime}}=A
\partial_{A1^{\prime}}
(x+y). \label{ej}
\end{equation}

\begin{pro}
Define
\begin{eqnarray*}
x(z)&=&-\exp(z^{00^{\prime}}-z^{01^{\prime}})-
\exp(z^{10^{\prime}}-z^{11^{\prime}}),\\
y(z)&=&\exp(z^{00^{\prime}}+z^{01^{\prime}})+
\exp(z^{10^{\prime}}+z^{11^{\prime}}).
\end{eqnarray*}
Then,
 the symmetry group of conformal
transformations for the SD connection (\ref{ej})
 is one--dimensional.
\end{pro}

{\bf Proof:} For the function ${\cal F}$ in (\ref{ecuab})
we find the expression
\[
{\cal F}=(\exp(z^{00^{\prime}}-z^{01^{\prime}})
+\exp(z^{00^{\prime}}+z^{01^{\prime}}))(\exp(z^{10^{\prime}}-z^{11^{\prime}})
+\exp(z^{10^{\prime}}+z^{11^{\prime}})).
\]
Upon substitution in (\ref{ecuab}) we obtain for $X$
the conditions
\begin{eqnarray*}
&&X^{A1^{\prime}}=0,\ A=0,1,\\
&&X^{00^{\prime}}+X^{10^{\prime}}=\mbox{div}_{\Omega}X,
\end{eqnarray*}
that follows from the rational character of the coefficients
$X^{AA^{\prime}}(z)$ of the vector field $X$.
 These conditions and the differential equations satisfied by any conformal
field $X$ imply that $X$ is proportional to
$\partial_{00^{\prime}}-\partial_{10^{\prime}}$. This
proves that
this SD connection obtained
 from the principal chiral field model
through the map prescribed by {\sc Theorem 2.2} have
at most a conformal symmetry.$\Box$

Analogous considerations show that  the same is true
for the KdV and NLS type equations. For if we let the zero--curvature
1--form be defined as
$\omega_+(\lambda)=U_1(dx+\lambda dt)$
for a constant element $U_1$ in {\got g} such that
 $\mbox{B}(U_1,U_1)=1$ and take
\begin{eqnarray*}
x(z)&=&-1/2(z^{01^{\prime}}(z^{00{\prime}})^2+
z^{11^{\prime}}
(z^{10{\prime}})^2)+
\exp(z^{00^{\prime}})+\exp(z^{10^{\prime}}),\\
t(z)&=&1/6((z^{00^{\prime}})^3+(z^{10^{\prime}})^3),
\end{eqnarray*}
we obtain as  conformal symmetry group, for the SD connection associated
to this solution of a NLS type equation, the one--dimensional
subgroup generated by
 \[
 \partial_{00^{\prime}}+\partial_{10^{\prime}}+
 z^{01^{\prime}}
\partial_{01^{\prime}}+
z^{11^{\prime}}\partial_{11^{\prime}}.
 \]
Therefore this SD connection associated to a NLS type equation
as in {\sc Theorem 2.1} has at most a conformal symmetry.

Finally, in the KdV case, for the constant vectors $U_0,U_1
\in\mbox{\got g}$
 satisfying $\mbox{B}(U_0,U_0)=\mbox{B}(U_1,U_1)=0$ and $\mbox{B}([U_0,U_1],
 [U_0,U_1])=1$, we define the zero--curvature 1--form
 \[
\omega_+(\lambda)=(\lambda U_1+U_0)(dx+\lambda dt)
\]
and the functions
 \begin{eqnarray*}
 t(z)&=&\exp(z^{00^{\prime}})+\exp(z^{10^{\prime}}), \\
 x(z)&=&-z^{01^{\prime}}\exp(z^{00^{\prime}})-z^{11^{\prime}}
 \exp(z^{10^{\prime}})
 +\exp(-z^{00^{\prime}})+\exp(3z^{10^{\prime}}).
 \end{eqnarray*}
The symmetry group of conformal transformations
for the associated
SD connection is generated in this case by
$\partial_{01^{\prime}}-\partial_{11^{\prime}}$.

 \end{document}